\documentstyle[11pt,paspconf,164page,164def,psfig]{article}
\begin{document}

\title{The Rosat$-$Green\,Bank Sample of Intermediate BL~Lac Objects}

\index{BL~Lac objects} 
\index{BL~Lac objects!samples}

\label{laure}

\markboth{Laurent-Muehleisen, Kollgaard, \& Feigelson}{The Rosat$-$Green\,Bank
BL~Lac Sample}

\author{S.\ A.\ Laurent-Muehleisen}
\affil{L-413, IGPP/LLNL, 7000 East Ave., Livermore, CA, 94550}

\author{R.\ I.\ Kollgaard}
\affil{MS 127, Fermi National Accelerator Lab, Batavia, IL, 60510}

\author{E.\ D.\ Feigelson}
\affil{Penn State Univ., Dept.\ of Astro.\ \& Astrophys., University Park,
PA, 16802}

\begin{abstract}
The Rosat$-$Green\,Bank BL\,Lac sample consists of 119 objects and smoothly
bridges the gap between the previously disparate subclasses of radio$-$ and
X-ray$-$selected objects.  Further study of this sample should provide
useful constraints to the unified scheme and help determine if modifications
are necessary.

\end{abstract}

\section{Introduction}
Prior to the launch of the ROSAT observatory, approximately 200 BL\,Lacs were
known, making them one of the rarest classes of AGN.  The vast majority of
these objects were discovered in either radio or X-ray surveys and belong to
two distinct observational subclasses: X-ray$-$ and radio$-$selected objects
\index{BL~Lac objects!X-ray selected} \index{BL~Lac objects!radio-selected}
(XBLs and RBLs respectively).  Of the two, RBLs exhibit more extreme properties
(e.g.\ Laurent-Muehleisen 1996, and references therein).  The data are largely
consistent with a unified scheme \index{unified scheme} which asserts that RBLs,
XBLs and FR\,I radio galaxies represent objects that are intrinsically the same,
but whose relativistic jets are oriented at increasingly larger angles to the
line-of-sight (Urry \& Padovani 1995; Kollgaard 1994).  Few objects with {\em
intermediate\/} \index{BL~Lac objects!intermediate objects} properties between
RBLs and XBLs are known, but their characterization is crucial for testing the
unified scheme.

\section{The RGB BL\,Lac Sample}
We have created such a sample of intermediate objects from a correlation of the
ROSAT All-Sky Survey (RASS) and the Green Bank 5\,GHz radio (GB) survey.  This
correlation consists of 2,127 objects for which we obtained followup VLA
observations which facilitated the unambiguous identification of optical
counterparts (Laurent-Muehleisen et al.\ 1997; Brinkmann et al.\ 1997).
Combining previously known BL\,Lacs with new objects we identified
spectroscopically, we produced the ROSAT$-$Green\,Bank (RGB) catalog of 119
BL\,Lacs (Laurent-Muehleisen 1996).  These BL\,Lacs span nearly three orders of
magnitude in both X-ray and radio flux (4$\times 10^{-13}$$<$F$_{\rm
x}$$<$4$\times 10^{-10}$\,erg\,s$^{-1}$cm$^{-2}$; 3$<$S$_{\rm r}$$<$
2160\,mJy), and have blue magnitudes which range from 13.3$<$B$<$21.2\,mag.  

\section{Discussion and Conclusions}
\begin{figure}[t]
\cl{\psfig{figure=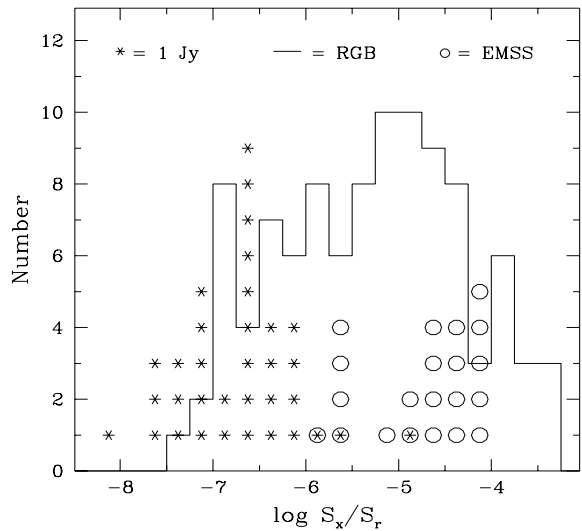,width=0.6\textwidth,height=2.89in}}
\caption{The ratio of the 2\,keV X-ray and 5\,GHz radio flux densities.  RBLs
are represented here by the 1\,Jy sample (Stickel, Fried, \& K\"{u}hr 1993) and
XBLs by the EMSS sample (Morris et al.\ 1991).  The RGB sample smoothly spans
the range from $-7.5$$<$$\log$ S$_{\rm x}$/S$_{\rm r}$$<$$-3.5$ and in
particular shows many objects with $\log$ S$_{\rm x}$/S$_{\rm r}$ between $-6$
and $-5$, a region which until now has been severely undersampled.\label{sxsr}}
\end{figure}

The division between RBLs and XBLs is usually based on the ratio of the X-ray
flux density to the radio flux density, where objects with S$_{\rm x}$/S$_{\rm
r}$$<$$-$5.5 are considered RBL-like.  The S$_{\rm x}$/S$_{\rm r}$ distribution
of the RGB BL\,Lacs shows this sample consists of both traditional XBL- and
RBL-like objects in addition to many objects with flux ratios that smoothly 
bridge the gap between them (Fig.~\ref{sxsr}).  

The RGB BL\,Lacs exhibit other properties which are consistent both with their
classification as intermediate objects and with the predictions of the unified
scheme.  However, similar to trends seen in other samples (e.g.\ Sambruna,
Maraschi, \& Urry 1996), the RGB BL\,Lacs exhibit some properties, particularly
in their high energy continuum, which indicate factors other than orientation
may be important.  Detailed study of a sample such as the RGB which contains
transitional objects should prove invaluable for testing the unified scheme and
characterizing any modifications that prove necessary.

\footnotesize\acknowledgments
This work was supported by NASA grant No.\ NAGW-2120 and the NASA Space Grant
Consortium through their Space Grant Fellow program.  \NRAOcredit


\begin{references}



\reference Brinkmann, W., et al. 1997. \aap, in press.

\reference Kollgaard, R.~I.\ 1994. {\em Vistas in Astronomy}, {\bf 38}, 29--75.

\reference Laurent-Muehleisen, S.~A., et al. 1996. PhD.\ Thesis, The
Pennsylvania State University.

\reference Laurent-Muehleisen, S.~A., et al. 1997. \aaps, {\bf 122}, 235--247.

\reference Morris, S.~L., et al. 1991. \apj, {\bf 380}, 49--65.

\reference Sambruna, R.~M., Maraschi, L., \& Urry, C.~M.\ 1996. \apj, {\bf 463},
444--465.

\reference Stickel, M., Fried, J., \& K\"{u}hr, H.\ 1993. \aaps, {\bf 98},
393--442.

\reference Urry, C.~M., \& Padovani. P. 1995. \pasp, {\bf 107}, 803--845.

\end{references}
\end{document}